\documentclass[10pt,aps,prl,twocolumn,superscriptaddress]{revtex4-1}

\usepackage{natbib}
\usepackage[dvips]{graphicx}
\usepackage{bm}
\usepackage{amsmath, amssymb}
\usepackage{upgreek}
\usepackage{url}

\newcommand{\BFA}{BaFe$_{2}$As$_{2}$}
\newcommand{\BFCA}{Ba(Fe$_{1-x}$Co$_{x}$)$_{2}$As$_{2}$}
\newcommand{\KFA}{KFe$_{2}$As$_{2}$}
\newcommand{\BFAP}{BaFe$_{2}$(As$_{1-x}$P$_{x}$)$_{2}$}

\newcommand{\BCA}{BaCo$_{2}$As$_{2}$}
\newcommand{\BFP}{BaFe$_{2}$P$_{2}$}
\newcommand{\BNA}{BaNi$_{2}$As$_{2}$}
\newcommand{\BNP}{BaNi$_{2}$P$_{2}$}

\newcommand{\Ts}{$T_{\mathrm{s}}$}
\newcommand{\Tc}{$T_{\mathrm{c}}$}
\newcommand{\cm}{cm$^{-1}$}
\newcommand{\Neff}{$N_{\mathrm{eff}}$}
\newcommand{\ND}{$N_{\mathrm{D}}$}

\newcommand{\LSCO}{La$_{2-x}$Sr$_{x}$CuO$_{4}$}

\begin{document}

\preprint{ver.\ 6.1}

\title{Strong electronic correlations in iron pnictides:\\ Comparison of the optical spectra for \BFA{}-related compounds}



\author{M.~Nakajima}
\email[]{m-nakajima@aist.go.jp}
\affiliation{Department of Physics, University of Tokyo, Tokyo 113-0033, Japan}
\affiliation{National Institute of Advanced Industrial Science and Technology, Tsukuba 305-8568, Japan}
\affiliation{JST, Transformative Research-Project on Iron Pnictides (TRIP), Tokyo 102-0075, Japan}
\author{S.~Ishida}
\affiliation{Department of Physics, University of Tokyo, Tokyo 113-0033, Japan}
\affiliation{National Institute of Advanced Industrial Science and Technology, Tsukuba 305-8568, Japan}
\affiliation{JST, Transformative Research-Project on Iron Pnictides (TRIP), Tokyo 102-0075, Japan}
\author{T.~Tanaka}
\affiliation{Department of Physics, University of Tokyo, Tokyo 113-0033, Japan}
\affiliation{JST, Transformative Research-Project on Iron Pnictides (TRIP), Tokyo 102-0075, Japan}
\author{K.~Kihou}
\affiliation{National Institute of Advanced Industrial Science and Technology, Tsukuba 305-8568, Japan}
\affiliation{JST, Transformative Research-Project on Iron Pnictides (TRIP), Tokyo 102-0075, Japan}
\author{Y.~Tomioka}
\affiliation{National Institute of Advanced Industrial Science and Technology, Tsukuba 305-8568, Japan}
\affiliation{JST, Transformative Research-Project on Iron Pnictides (TRIP), Tokyo 102-0075, Japan}
\author{T.~Saito}
\affiliation{Department of Physics, Chiba University, Chiba 263-8522, Japan}
\author{C.~H.~Lee}
\affiliation{National Institute of Advanced Industrial Science and Technology, Tsukuba 305-8568, Japan}
\affiliation{JST, Transformative Research-Project on Iron Pnictides (TRIP), Tokyo 102-0075, Japan}
\author{H.~Fukazawa}
\affiliation{JST, Transformative Research-Project on Iron Pnictides (TRIP), Tokyo 102-0075, Japan}
\affiliation{Department of Physics, Chiba University, Chiba 263-8522, Japan}
\author{Y.~Kohori}
\affiliation{JST, Transformative Research-Project on Iron Pnictides (TRIP), Tokyo 102-0075, Japan}
\affiliation{Department of Physics, Chiba University, Chiba 263-8522, Japan}
\author{T.~Kakeshita}
\affiliation{Department of Physics, University of Tokyo, Tokyo 113-0033, Japan}
\affiliation{JST, Transformative Research-Project on Iron Pnictides (TRIP), Tokyo 102-0075, Japan}
\author{A.~Iyo}
\affiliation{National Institute of Advanced Industrial Science and Technology, Tsukuba 305-8568, Japan}
\affiliation{JST, Transformative Research-Project on Iron Pnictides (TRIP), Tokyo 102-0075, Japan}
\author{T.~Ito}
\affiliation{National Institute of Advanced Industrial Science and Technology, Tsukuba 305-8568, Japan}
\affiliation{JST, Transformative Research-Project on Iron Pnictides (TRIP), Tokyo 102-0075, Japan}
\author{H.~Eisaki}
\affiliation{National Institute of Advanced Industrial Science and Technology, Tsukuba 305-8568, Japan}
\affiliation{JST, Transformative Research-Project on Iron Pnictides (TRIP), Tokyo 102-0075, Japan}
\author{S.~Uchida}
\affiliation{Department of Physics, University of Tokyo, Tokyo 113-0033, Japan}
\affiliation{JST, Transformative Research-Project on Iron Pnictides (TRIP), Tokyo 102-0075, Japan}


\date{\today}

\begin{abstract}

We carried out combined transport and optical measurements for \BFA{} and five isostructural transition-metal (\textit{TM}) pnictides. The low-energy optical conductivity spectra of these compounds are, to a good approximation, decomposed into a narrow Drude (coherent) component and an incoherent component. The iron arsenides, \BFA{} and \KFA{}, are distinct from other pnictides in their highly incoherent charge dynamics or bad metallic behavior with the coherent Drude component occupying a tiny fraction of the low-energy spectral weight. The fraction of the coherent spectral weight or the degree of coherence is shown to be well correlated with the \textit{TM}-pnictogen bond angle and the electron filling of \textit{TM} 3d orbitals, which are measures of the strength of electronic correlations. The iron arsenides are thus strongly correlated systems, and the doping into \BFA{} controls the strength of electronic correlations. This naturally explains a remarkable asymmetry in the charge dynamics of electron- and hole-doped systems, and the unconventional superconductivity appears to emerge when the correlations are fairly strong.

\end{abstract}


\maketitle

The superconducting (SC) phase with unconventional pairing symmetry and with fairly high transition temperature (\Tc{}) in iron-based superconductors emerges by doping (chemical substitution) into a parent compound showing an antiferromagnetic-orthorhombic (AFO) order \cite{Johnston2010}. This is similar to the high-\Tc{} cuprates, in which the parent compounds are Mott insulators due to strong electronic correlations. In contrast, the parent compounds of the iron-based superconductors are metals, albeit not good metals, and are not close to a Mott insulating state. From this, the iron pnictides/chalcogenides are usually thought to be in the intermediate correlation regime. Nevertheless, they possess features of strong electronic correlations, such as frozen spin magnetic moments \cite{Cruz2008,Huang2008} and temperature ($T$)-linear resistivity, characteristic of a non-Fermi liquid \cite{Leyraud2009,Kasahara2010}. These ``bad metallic'' behaviors are usually observed for metals in close proximity to a Mott insulator and raise a puzzling question on the role of electronic correlations in the iron-based compounds, specifically iron arsenides.

There exist theoretical attempts to solve this puzzle by incorporating the effect of the Hund's rule coupling ($J_{\mathrm{H}}$) in addition to the onsite Coulomb interaction ($U$), suggesting that, even when both $J_{\mathrm{H}}$ and $U$ are not large enough, the former coupling may be responsible for the aspects of strong correlations in multi-orbital metals \cite{Yin2011,Georges2013}. However, so far there have been few experimental studies that are relevant to this issue \cite{Qazilbash2009}, and it is far from clear why the iron arsenides are so special.

Among various families of iron-based superconductors, the family with a ThCr$_2$Si$_2$ crystal structure (so-called 122 system), representatively \BFA{} and its doped compounds, has been most extensively studied. An advantage of this family is that isostructural compounds with various transition-metal (\textit{TM}) and pnictogen (\textit{Pn}) atoms can be synthesized, including the end members, e.g., \BCA{}, accessible via doping from \BFA{} going through the SC regime. Therefore, this 122 system provides an opportunity to experimentally address the key questions: why the Fe-As combination is unique, how the effects of electronic correlations vary among each member of this family, and what is a major control parameter for superconductivity to emerge by doping.

In the present work, the optical spectra of \BFA{} and its five isostructural compounds (\BCA{}, \BFP{}, \KFA{}, \BNA{}, and \BNP{}) are investigated in the high-temperature paramagnetic-tetragonal (PT) phase. We show that the degree of incoherence in the charge dynamics is well correlated with the strength of electronic correlations and that strong correlations make the iron arsenides distinct from others. The result also indicates that doping controls the strength of correlations via an application of chemical pressure and a change in the electron filling of the \textit{TM} 3d orbitals.


Single crystals of \BFA{}, \BCA{}, \BFP{}, \KFA{}, and \BNA{} were grown by the self-flux method \cite{Nakajima2010,Nakajima2012a,Kihou2010}. The undoped parent compound \BFA{}, the quality of which is remarkably improved by annealing, shows very low residual resistivity ($<$ 10 $\upmu\Omega$\,cm) at low temperatures in the AFO phase, and the AFO-PT transition temperature \Ts{} is 143 K, the highest among so far reported \cite{Nakajima2011,Ishida2011}. For \BNP{}, we used a Sn-flux method to obtain sizable single crystals. Resistivity measurements were performed on the \textit{ab} plane using the four-terminal method. Optical reflectivity was measured at $T=300$ K with the incident light almost normal to the \textit{ab} plane in the frequency range of 50--40000 \cm{} using a Fourier transform infrared spectrometer (Bruker IFS113v) and a grating monochromator (JASCO CT-25C). The optical conductivity was derived from the Kramers-Kronig transformation of the reflectivity spectrum. The Hagen-Rubens or Drude-Lorentz formula was used for the low-energy extrapolation in order to smoothly connect to the spectrum in the measured region and to fit the measured resistivity value at $\omega=0$.


\begin{figure}
\includegraphics[width=70mm,clip]{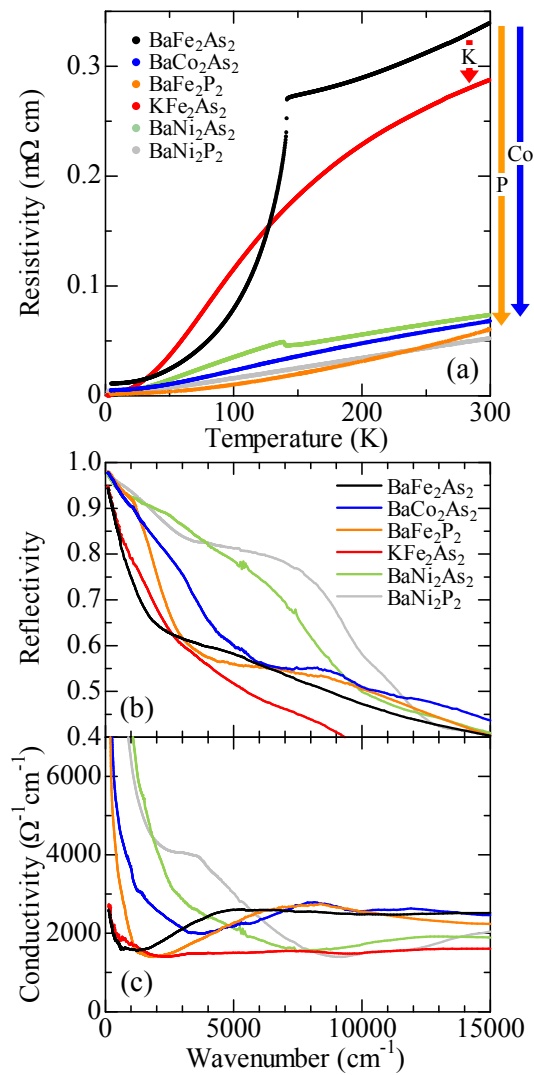}%
\caption{(a) Temperature dependence of the in-plane resistivity for \BFA{}, \BCA{}, \BFP{}, \KFA{}, \BNA{}, and \BNP{}. (b) Reflectivity and (c) optical conductivity spectrum for each compound measured at $T=300$ K.}
\end{figure}

Figure 1(a) displays $\rho(T)$ for all the compounds of \textit{TM} pnictides investigated in the present work. Anomalies observed at 143 K for \BFA{} and at 130 K for \BNA{} correspond to the AFO-PT and the triclinic-tetragonal structural transition \cite{Sefat2009a}, respectively. At high temperatures, \BFA{} and \KFA{} exhibit a bad metallic behavior with relatively high resistivity and its weak temperature dependence. The magnitudes of resistivity for other pnictides are considerably low, and the resistivity shows either $T$ or $T^2$ dependence, typical of metals. On the other hand, all the compounds, even \BFA{} and \KFA{}, show very low resistivity of $\sim$ 10 $\upmu\Omega$\,cm or lower at low temperatures. Focusing on the resistivity in the high-temperature region, one finds that these pnictides are classified into two distinct groups: bad metals with relatively high resistivity $\sim$ 300 $\upmu\Omega$\,cm at 300 K (\BFA{} and \KFA{}) and good metals with low resistivity $\sim$ 60 $\upmu\Omega$\,cm at 300 K (\BCA{}, \BFP{}, and the two nickel pnictides). Co and P doping turn out to bridge these two classes, while the system remains a bad metal for K doping.



The reflectivity spectra at $T=300$ K for \BFA{} and the related materials are shown in Fig.\ 1(b). The spectrum of \BFA{} shows a plasma edge at $\sim$ 2000 \cm{}. The plasma edge is less clear in comparison with typical metals such as gold or silver, in which an edge definitely separates the energy regions showing high and low reflectivity. The plasma-edge frequency for \BFA{} is the lowest among the six, shifting to higher frequencies in the order of \KFA{}, \BFP{}, \BCA{}, \BNA{}, and \BNP{}, suggesting that the carrier density forming the plasma edge (carrier effective mass) increases (decreases) in this order.

Figure 1(c) shows the optical conductivity spectra of the six compounds. The spectrum of \BFA{} has a small peak at $\omega=0$ and a long tail. It seems that a tiny Drude component is embedded in a large incoherent background. The low-energy spectrum of \KFA{} exhibits basically the same feature. In contrast, for the other four compounds, a peak centered at $\omega=0$ is much higher than that for the two iron arsenides, consistent with the lower resistivity. In particular, the two nickel-based compounds show a Drude component with significantly larger spectral weight.


\begin{figure}
\includegraphics[width=65mm,clip]{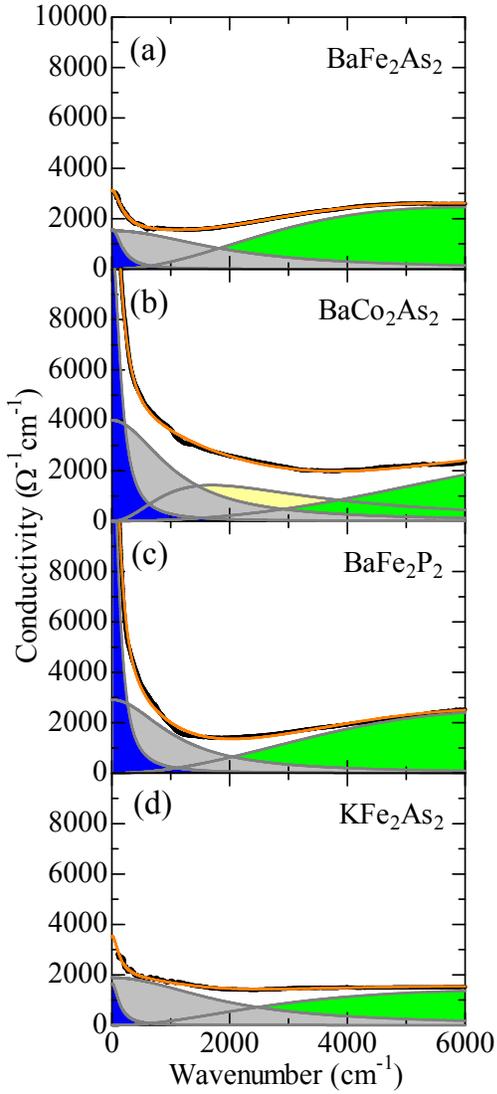}%
\caption{Decomposition of the conductivity spectrum for the end members of the present three systems. Black and orange curves indicate the experimental data and the fitting results, respectively.}
\end{figure}

To quantitatively analyze charge dynamics in these compounds, we decompose the conductivity spectrum into several components \cite{Nakajima2010,Wu2010}. The results for the representative end compounds for the three types of doping (\BFA{}, \BCA{}, \BFP{}, and \KFA{}) are displayed in Figs.\ 2(a)--2(d). The black and the orange lines are the experimental data and the fitting results, respectively. For all the four compounds, it turns out that the spectrum in the low-energy region is dominated by two distinct components, narrow and broad Drude components, shaded in blue and gray, respectively. A higher-energy component approximated by a Lorentzian function (in green) corresponds to an interband excitation. In the spectrum of \BFA{}, the weight of the narrow Drude term is quite small, and the other Drude component is extremely broad with its width of 2000 \cm{} or larger, indicative of incoherent character \cite{Nakajima2010}. The dominance of the broad Drude component is also seen in the case of \KFA{} [Fig.\ 2(d)]. Such a dominant contribution of the broad Drude component is characteristic of iron-arsenide compounds and is responsible for the high resistivity at high temperatures. By contrast, for \BCA{} and \BFP{} [Figs.\ 2(b) and 2(c), respectively], the narrow Drude component possesses much larger spectral weight than those for \BFA{} and \KFA{}, and even the broad Drude term is significantly narrowed in width, indicating that most of carriers gain coherence.

\begin{figure}
\includegraphics[width=65mm,clip]{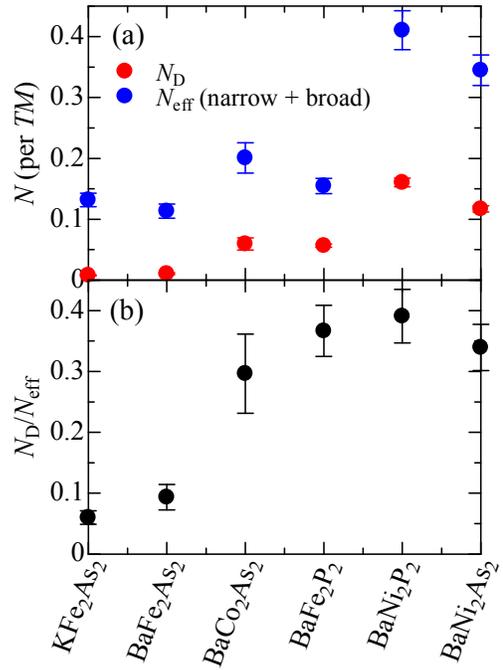}%
\caption{(a) Weight of the narrow Drude component (\ND{}) and the total weight of the narrow and broad Drude components (\Neff{}), and (b) the ratio of \ND{} to \Neff{}, as a measure of coherence for various \textit{TM} pnictides. The compounds are arranged in the order of the \textit{Pn}-\textit{TM}-\textit{Pn} angle $\alpha_{\mathrm{bond}}$ (121.7$^{\circ}$ for \BNP{} \cite{Keimes1997} and 122.9$^{\circ}$ for \BNA{} \cite{Sefat2009a}).}
\end{figure}

The values of the weight of the narrow Drude component \ND{} ($= \frac{m_0}{m^*} n_{\mathrm{D}}$; $m_0$, $m^*$, and $n_{\mathrm{D}}$ being the free electron mass, the carrier effective mass, and the coherent carrier density, respectively) for the present six end pnictides are plotted in Fig.\ 3(a). \ND{} of \KFA{} is comparable with that of \BFA{}, and the values for \BCA{} and \BFP{} is larger by a factor of about 6 compared with that for \BFA{}. The fitting of the spectra of the two nickel pnictides using the two Drude components is as good as those for \BCA{} and \BFP{} (see Supplementary information), although the weight of the narrow Drude component is significantly larger. In Fig.\ 3(a), also plotted is the low-energy spectral weight \Neff{}, contributed by both coherent and incoherent carrier dynamics. Since it is difficult to unambiguously separate the incoherent weight from that for the interband excitations, we estimate \Neff{} as a sum of the weight of the narrow and the broad Drude component obtained by the present fitting \cite{Neff}. The fraction of the coherent Drude weight \ND{}/\Neff{}, a measure of the degree of coherence, is shown in Fig.\ 3(b). The coherent fraction of \BFA{} and \KFA{} is distinctively low as compared with that for other \textit{TM} pnictides.

\begin{figure}
\includegraphics[width=75mm,clip]{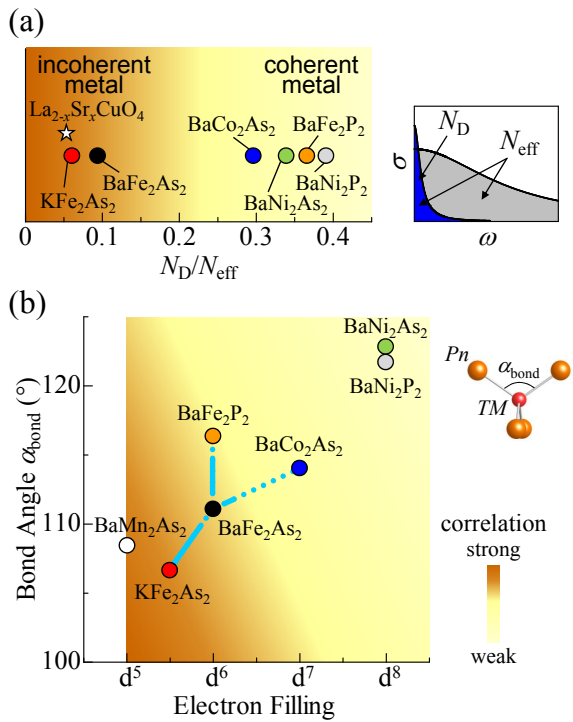}%
\caption{(a) Fraction of the coherent Drude spectral weight \ND{}/\Neff{} for various \textit{TM} pnictides with the same crystal structure, including one typical underdoped cuprate. (b) \textit{TM} pnictides mapped on the diagram of \textit{Pn}-\textit{TM}-\textit{Pn} bond angle $\alpha_{\mathrm{bond}}$ vs electron filling of \textit{TM} d orbitals in the \textit{TM}$^{2+}$ state. Blue dotted and solid lines are trajectories of three types of doping into \BFA{}, and superconductivity is observed in the regions indicated by the solid blue lines. Smaller $\alpha_{\mathrm{bond}}$ and lower electron filling toward d$^5$, just half filling, point toward stronger electronic correlation. Correlation strength is schematically scaled with indicated color code.}
\end{figure}

Qazilbash \textit{et al.}\ estimated the strength of electronic correlation of LaFePO from a ratio between experimental plasma frequency of conduction electrons (holes) and that estimated from the band-structure calculation in the LDA \cite{Qazilbash2009}. They show that the ratio for LaFePO ($\sim$ 0.5) is larger than that for \BFA{} ($\sim$ 0.3), indicating that an enhancement of the effective mass due to correlation in the arsenides is larger than that in the phosphides. The present \ND{}/\Neff{} may be regarded as an alternative measure of the electronic correlations, which is more appropriate for a quantitative comparison of the correlation strength among \textit{TM} pnictides. The values of \ND{}/\Neff{} are replotted on the diagram shown in Fig.\ 4(a), which should be compared with the ``mass-renormalization'' diagram by Qazilbash \textit{et al.} \cite{Qazilbash2009}. The values of \ND{}/\Neff{} for the iron arsenides are distinctively smaller than those for other \textit{TM} pnictides. They are comparable with those for the hole-underdoped cuprates (\textit{e.g.} \LSCO{} with $x=0.10$) assuming the in-plane spectrum decomposed into coherent (simple Drude) and incoherent (so-called mid-infrared) components \cite{Uchida1991}. In view of the strong incoherence in the cuprates arising from strong electronic correlation which frustrates coherent carrier motion, it is likely that electronic correlations are also a source of incoherence. The on-site Coulomb interaction ($U$) in the high-\Tc{} cuprates is considerably stronger than that in \BFA{} \cite{Miyake2010b}. Hence, \ND{}/\Neff{} likely measures electronic correlations arising also from the Hund's coupling ($J_{\mathrm{H}}$) reflecting the multi-orbital nature of the iron-based materials \cite{Yin2011,Misawa2012}, and in this sense, the iron arsenides might be a strongly correlated system. 

It is well known that the electronic structure of the iron pnictides is sensitive to a change in local lattice parameters, particularly to the shape of a FeAs$_4$ tetrahedron usually parameterized by the As-Fe-As bond angle $\alpha_{\mathrm{bond}}$ \cite{Miyake2010a,Usui2011}. As $\alpha_{\mathrm{bond}}$ increases, the hybridization of Fe 3d and As 4p orbitals becomes stronger. This leads to an increase in bandwidth ($W$) and hence to weakening of electron correlations (either $U/W$ or $J_{\mathrm{H}}/W$). $\alpha_{\mathrm{bond}}$ for \BFP{} (116.4$^{\circ}$) is larger than that for \BFA{} (111.1$^{\circ}$) \cite{Rotter2010}. Therefore, it is expected that the P doping into \BFA{} works as a chemical pressure and weakens electronic correlations. This appears to be supported by the result of quantum oscillations for \BFAP{}, which shows that $m^*$ decreases by a factor of 2--3 from $x=0.3$ to 1 \cite{Shishido2010}. The increase in $m^*$ in this system was ascribed to a quantum critical behavior toward a critical point near $x=0.3$, and a subsequent decrease in $m^*$ below $x=0.3$ was suggested by the penetration-depth measurement in the SC state \cite{Hashimoto2012}. However, in the normal state well above \Tc{}, the variation of \ND{} or resistivity with $x$ is monotonous, showing no trace of the quantum critical behavior \cite{Nakajima2013} (see Fig.\ S2 in Supplementary information). Thus, a major effect of the isovalent P doping is likely to weaken electronic correlations \cite{Nakajima2012b}. This would be also the case with the Co doping, in which $\alpha_{\mathrm{bond}}$ for \BCA{} (114.0$^{\circ}$ \cite{Sefat2009b}) is significantly larger than that for \BFA{} (but smaller than that for \BFP{}). Although there is no reported data that clearly indicate a decrease in $m^*$ in \BFCA{}, a possible decrease in $m^*$ due to the chemical-pressure effect may also contribute to the increase in \ND{} in addition to the increase in the coherent electron density.

To the contrary, $\alpha_{\mathrm{bond}}$ for \KFA{} (106.7$^{\circ}$ \cite{Rotter2008}) is smallest among the \textit{TM} pnictides investigated here. Thus, it is likely that the effect of K doping into \BFA{} is to apply a negative chemical pressure, which would make electronic correlation stronger. In fact, an enhancement of $m^*$ by a factor of 4--5 is observed by quantum oscillations \cite{Terashima2010,Terashima2011}. 
A trend that a compound with smaller $\alpha_{\mathrm{bond}}$ has larger effective mass $m^*$ is in agreement with the trend found theoretically by Yin \textit{et al.} \cite{Yin2011} between bond angle and strength of the Hund's coupling. Here, we also found a correlation between $\alpha_{\mathrm{bond}}$ and \ND{}/\Neff{}. The Hund's correlation tends to inhibit mixing of the orbitals, which would frustrate the electron motion.

Distinct \ND{}/\Neff{} values for \BCA{} and \KFA{} and the striking asymmetry in the evolution of the in-plane resistivity with electron and hole doping (see Supplementary information) suggest that, in addition to the chemical-pressure effect, an increase in electron density or electron filling may also affects the strength of the electronic correlation. The hole (K) doping reduces the Fe 3d orbital filling from d$^6$ to d$^{5.5}$, whereas the electron (Co) doping increases it toward d$^7$. d$^5$ is just half filling, at which the electronic correlation is maximal, making the compound (such as BaMn$_2$As$_2$ \cite{Singh2009}) a Mott insulator. Hence, the hole doping into \BFA{} or the reduction in the electron filling from d$^6$ points toward stronger correlation \cite{Misawa2012,Medici2009}, explaining the remarkable asymmetry between the evolutions of the resistivity with Co and K doping \cite{Nakajima2013}.

In Fig.\ 4(b), various pnictides are mapped on the ``electronic-correlation'' diagram of $\alpha_{\mathrm{bond}}$ (bandwidth) vs electron filling. Smaller $\alpha_{\mathrm{bond}}$ and lower filling point toward stronger correlation. One sees that the correlation in \KFA{} is strongest among the six compounds in the present study, whereas that in the nickel pnictides are weakest, and that the material dependence of the correlation strength is in overall agreement with that of the degree of incoherence shown in Fig.\ 4(a). Also, clear is that a major role of doping into \BFA{} is to adjust correlation, making it weaker (stronger) for P or Co doping (K doping), as indicated by the dashed lines, consistent with the trend in \ND{}/\Neff{} as well as that in the reported effective mass.



In summary, by studying the transport and optical properties of various \textit{TM} pnictides with the same structure, we have demonstrated that iron arsenides, \BFA{} and \KFA{}, are distinct from other \textit{TM} pnictides and characterized by highly incoherent charge dynamics. Mapping these compounds on the electronic correlation diagram reveals that the degree of incoherence correlates well with the strength of electronic correlations. The electronic correlations are stronger in iron arsenides, and doping, which links \BFA{} to \BCA{} or to \BFP{}, enhances coherence by weakening the electronic correlations. On the other hand, the link to \KFA{} reduces coherence by enhancing correlations. It is inferred that superconductivity in pnictides (and possibly in chalcogenides) emerges in the doping regions, in which the electronic correlations are fairly strong as indicated by the solid blue lines on the links connecting \BFA{} to the three compounds, and either too weak or too strong correlations are not favorable for high-\Tc{} superconductivity. This condition is fulfilled in undoped \BFA{}. Its ground state, however, is a metal with the AFO order, but it readily transforms to high-\Tc{} superconductors once the AFO order is suppressed by doping or pressure.


\begin{acknowledgments}

M.N. and S.I. thank the Japan Society for the Promotion of Science (JSPS) for the financial support. Discussions with T. Misawa, M. Imada, and A. Fujimori were helpful in preparing this manuscript. This work was supported by the Japan-China-Korea A3 Foresight Program and Grant-in-Aid for JSPS Fellows from JSPS, Grant-in-Aid for Scientific Research from JSPS and the Ministry of Education, Culture, Sports, Science, and Technology, Japan, and the Strategic International Collaborative Research Program (SICORP) from Japan Science and Technology Agency.

\end{acknowledgments}

\end{document}